# Numerical Investigation of Mach 10 Kerosene-Fueled Oblique Detonation Waves at Different Flight Altitudes

Yunfeng Liu[1,2]
1. Institute of Mechanics, Chinese Academy of Sciences, Beijing 100190, China
2. School of Engineering Science, University of Chinese Academy of Sciences, Beijing 100049, China
liuyunfeng@imech.ac.cn

**Abstract:** The initiation and propagation mechanisms of kerosene-fueled oblique detonation waves are key issues in the development of oblique detonation wave engines. In this study, numerical simulations of kerosene-fueled oblique detonation at a flight Mach number of 10 were conducted using the two-dimensional conservative Euler equations and a second-order two-step global chemical reaction model. The objective was to compare wedge-induced initiation with bump-forced initiation. The results show that, at different flight altitudes, the oblique detonation flow field follows the ρL binary scaling law. Wedge-induced initiation requires highly precise matching of parameters such as the inlet-exit conditions, wedge angle and length, and equivalence ratio. In contrast, bump-forced initiation exploits the high total temperature and total pressure at the stagnation point and therefore does not require precise matching of these parameters. Therefore, the bump-forced initiation method provides reliable oblique detonation initiation and a stable oblique detonation flow field.

## 1. Introduction

Oblique detonation wave engines offer high thermodynamic efficiency, rapid heat release, high specific impulse, and simple structure. As a new generation of air-breathing hypersonic propulsion technology, they have attracted extensive attention worldwide [1-4]. Zhang et al. conducted the first full-scale free-jet experiment of a hydrogen-fueled oblique detonation wave engine at a simulated flight Mach number of 9 in a shock tunnel, thereby demonstrating the feasibility of the concept [5]. The University of Central Florida has conducted direct-connect experiments on hydrogen-fueled oblique detonation and developed a Mach 10 flight-test program for an oblique detonation wave engine [6-9].

Compared with hydrogen, kerosene has a higher volumetric energy density and is easier to store, making it more attractive for engineering applications. To address the long ignition delay time and initiation difficulty of kerosene, Han et al. proposed a bump-forced initiation technique and successfully conducted a shock-tunnel free-jet experiment at a simulated flight Mach number of 9. Their results demonstrated the engineering feasibility of a liquid kerosene-fueled oblique detonation wave engine [10,11]. Zhang et al. used bump-induced initiation technique in a shock-tunnel free-jet experiment on a kerosene-fueled oblique detonation wave engine at a simulated flight Mach number of 10 [12]. Zhao et al. also conducted a direct-connect experiment on kerosene-fueled oblique detonation with bump-induced initiation technique at a simulated flight Mach number of 8 [13]. Three-dimensional numerical simulations of a full-scale kerosene-fueled oblique detonation wave engine showed that the engine could operate stably over the Mach 8-10 range [14,15]. Yuan and Han numerically investigated the flow characteristics of a kerosene-fueled oblique detonation wave engine [16].

Research on kerosene-fueled oblique detonation faces several challenges. It has very long ignition delay time, making oblique detonation waves difficult to establish. The chemical kinetics of liquid kerosene are complex, resulting in high computational costs. In addition, kerosene-droplet evaporation and breakup strongly affect the initiation length and wave-front morphology. Using a Eulerian-Lagrangian method, Ren et al. investigated gas-liquid two-phase oblique detonation and clarified the effects of kerosene-droplet evaporation and initial droplet diameter on oblique detonation initiation and stability [17]. Numerical simulations by Tian et al. showed that temperature gradients caused by kerosene-droplet evaporation and nonuniform reactant distributions destabilize detonation waves [18]. Zhang et al. examined the kerosene-droplet evaporation models and found that droplet evaporation reduces the heat release rate and suppresses wave-front instability [19]. Wang et al. showed that different kerosene-droplet breakup models significantly affect oblique detonation wave structures [20]. Sun et al. investigated the effects of kerosene pre-injection on oblique detonation initiation and flow field stability [21].

In summary, the initiation and propagation mechanisms of kerosene-fueled oblique detonation waves constitute an important fundamental problem, but has received relatively limited attention. Because kerosene has a long ignition delay time and is difficult to initiate, conventional wedge-induced initiation becomes less reliable, making bump-forced initiation necessary. This study numerically investigates the initiation and propagation mechanisms of gaseous kerosene-air oblique detonation waves at a flight Mach number of 10 and at different flight altitudes. The objective is to compare wedge-induced initiation with bump-forced initiation and to provide a guidance for the engineering application of kerosene-fueled oblique detonation wave engines.

## 2. Governing Equations and Physical Model

The governing equations consist of the two-dimensional conservative Euler equations coupled with a second-order two-step global chemical reaction model [22]. Molecular vibrational energy excitation is considered, and the gas is assumed to be in thermal equilibrium. This second-order two-step reaction model is consistent with collision theory and is of pressure-dependence. A global chemical reaction model must reproduce both the ignition delay time and the heat release rate (HRR); therefore, a good global model at least contains two steps, one for ignition delay time and the other for heat release rate.

$$\frac{\partial \rho}{\partial t}+\frac{\partial \rho u}{\partial x}+\frac{\partial \rho v}{\partial y}=0 \tag{1}$$

$$\frac{\partial \rho u}{\partial t}+\frac{\partial\left(\rho u^2+p\right)}{\partial x}+\frac{\partial \rho u v}{\partial y}=0 \tag{2}$$

$$\frac{\partial \rho v}{\partial t}+\frac{\partial \rho u v}{\partial x}+\frac{\partial\left(\rho v^2+p\right)}{\partial y}=0 \tag{3}$$

$$\frac{\partial \rho e}{\partial t}+\frac{\partial\left(\rho e+p\right)u}{\partial x}+\frac{\partial\left(\rho e+p\right)v}{\partial y}=0 \tag{4}$$

$$\frac{\partial \rho\alpha}{\partial t}+\frac{\partial \rho u\alpha}{\partial x}+\frac{\partial \rho v\alpha}{\partial y}=\dot{\omega}_{\alpha} \tag{5}$$

$$\frac{\partial \rho\beta}{\partial t}+\frac{\partial \rho u\beta}{\partial x}+\frac{\partial \rho v\beta}{\partial y}=\dot{\omega}_{\beta} \tag{6}$$

$$e=\frac{p}{\left(\gamma-1\right)\rho}+\frac{1}{2}\left(u^{2}+v^{2}\right)+\beta q+e_{v} \tag{7}$$

$$e_{v}=\frac{\theta_{v}R}{\exp\left(\theta_{v}/T-1\right)} \tag{8}$$

$$\dot{\omega}_{\alpha}=-A_{I}\rho^{2}\exp\left(-\frac{Ea_{I}}{RT}\right) \tag{9}$$

$$\dot{\omega}_{\beta}=\begin{cases}0\ (if\ \ \alpha>0)\\ -A_{R}\rho^{2}\beta^{2}\exp\left(-\frac{Ea_{R}}{RT}\right)\ \left(if\ \ \alpha\leq 0\right)\end{cases} \tag{10}$$

$$\gamma\left(\beta\right)=\frac{\gamma_{1}R_{1}\beta/(\gamma_{1}-1)+\gamma_{2}R_{2}\left(1-\beta\right)/(\gamma_{2}-1)}{R_{1}\beta/(\gamma_{1}-1)+R_{2}\left(1-\beta\right)/(\gamma_{2}-1)} \tag{11}$$

$$R\left(\beta\right)=R_{1}\beta+R_{2}\left(1-\beta\right) \tag{12}$$

In the governing equations, $\rho$, $p$, $T$, $u$, $v$, $e$, $e_v$, $\theta_v$, $q$, $\gamma$, and $R$ denote density, pressure, temperature, velocity components, total internal energy per unit mass, vibrational energy per unit mass, characteristic vibrational temperature, heat release per unit mass, specific heat ratio, and gas constant, respectively. In the second-order two-step global chemical reaction model, $\alpha$ and $\beta$ stand for the reactant mass fractions for the induction and heat release reactions, respectively. $\dot{\omega}_{\alpha}$ and $\dot{\omega}_{\beta}$ stand for the source terms, while $A_I$, $A_R$, $Ea_I$, and $Ea_R$ are the pre-exponential factors and activation energies. The specific heat ratio and gas constant are functions of β; the subscripts in Eqs. (11) and (12) denote the reactants and products, respectively.

The parameters above are determined as follows. The specific heat ratio, gas constant, and Chapman-Jouguet (CJ) detonation velocity $D_{\mathrm{CJ}}$ are calculated using the Gaseq software. The heat release per unit mass $q$ is calculated from $D_{\mathrm{CJ}}=\sqrt{2\left(\gamma_{2}^{2}-1\right)q}$.

The activation energies $Ea_I$ and $Ea_R$ are calibrated from the slopes of the curves of ignition delay time. Finally, the pre-exponential factors $A_I$ and $A_R$ are calibrated against the ignition delay time.

For the stoichiometric gaseous kerosene-air mixture, the parameters are $\alpha_1 = 1.0$, $\beta_1 = 1.0$, $\beta_2 = 0$, $\gamma_1 = 1.35$, $\gamma_2 = 1.25$, $R_1 = 296\ \mathrm{J/(kg \cdot K)}$, $R_2 = 278\ \mathrm{J/(kg \cdot K)}$, $q = 2.5\ \mathrm{MJ/kg}$, $A_I = 2.5\times10^9\ \mathrm{kg^{-1}s^{-1}m^3}$, $A_R = 1.5\times10^9\ \mathrm{kg^{-1}s^{-1}m^3}$, $Ea_I = 5.13\ \mathrm{MJ/kg}$, and $Ea_R = 5.07\ \mathrm{MJ/kg}$. For the Mach 10 flight conditions, nitrogen is the dominant species of both the reactants and products; therefore, the characteristic vibrational temperature of nitrogen, $\theta_v = 3390\ \mathrm{K}$, is used for the mixture. Figure 1 compares the ignition delay time predicted by the present model with experimental data and the detailed reaction model, which shows that the agreement is good. The experimental and detailed model ignition delay time are reported in Ref. [11].

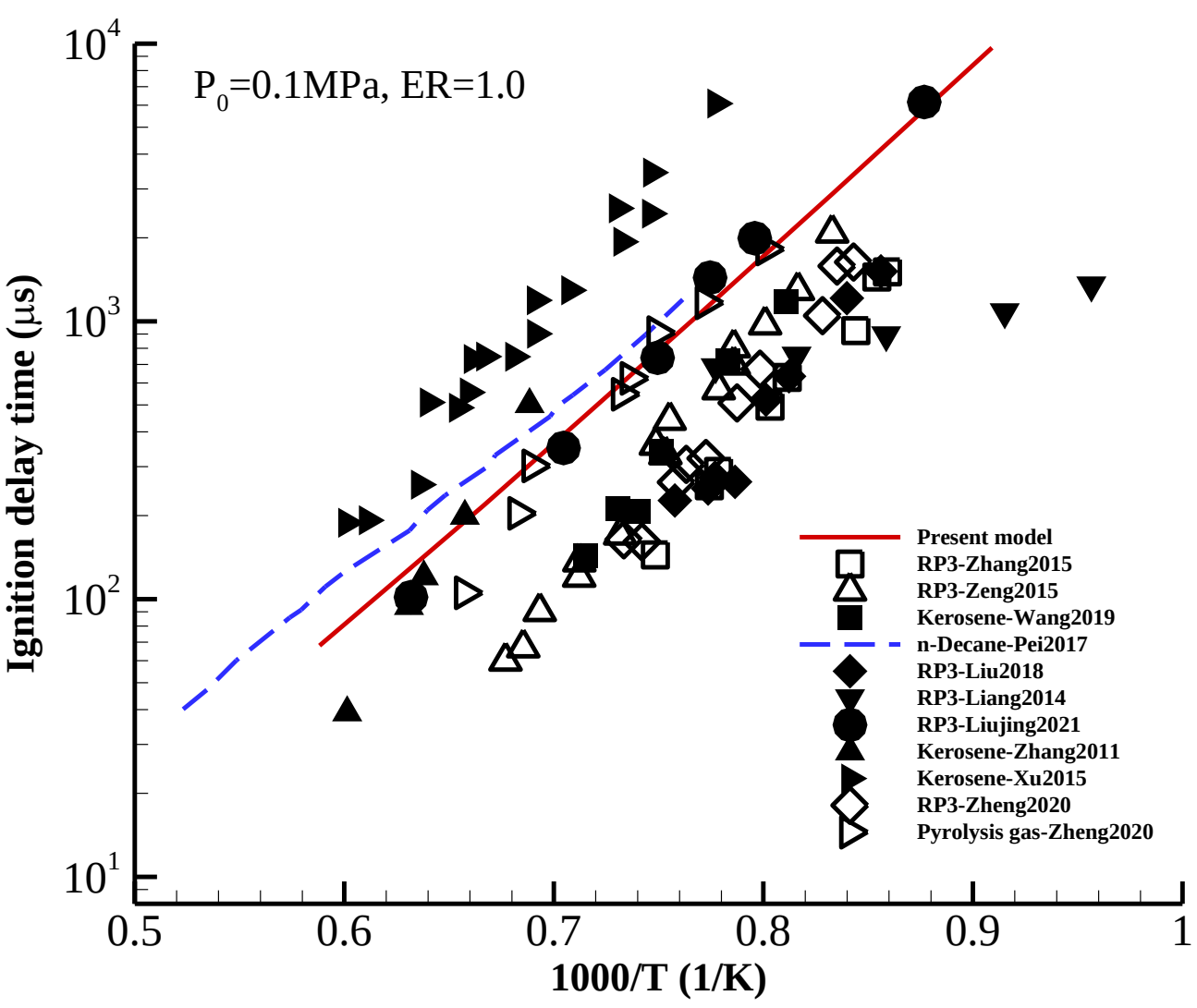

Fig. 1. Comparison of ignition delay times

In the homemade code, the numerical scheme employs a second-order essentially non-oscillatory (ENO) method and a third-order total-variation-diminishing (TVD) Runge-Kutta method. Both wedge-induced initiation and bump-forced initiation are simulated, and the corresponding computational domains are shown in Fig. 2. The computational domain is 40 mm long and 24 mm wide, with a uniform grid spacing of 20 μm. The Courant–Friedrichs–Lewy (CFL) number is 0.31. The wedge is 40 mm long and has an angle of 20° , as shown in Fig. 2(a), being comparable to the wedge dimensions of a full-scale oblique detonation wave engine [12]. For bump-forced initiation, as shown in Fig. 2(b), the bump is a square with a side length of 0.6 mm. It is located 10 mm downstream of the wedge tip, while the wedge tip is 2 mm from the left boundary. Freestream boundary conditions are imposed at the left and upper boundaries; outlet conditions are imposed at the right boundary and along the boundary upstream of the wedge tip; and a slip wall condition is applied to the solid surface.

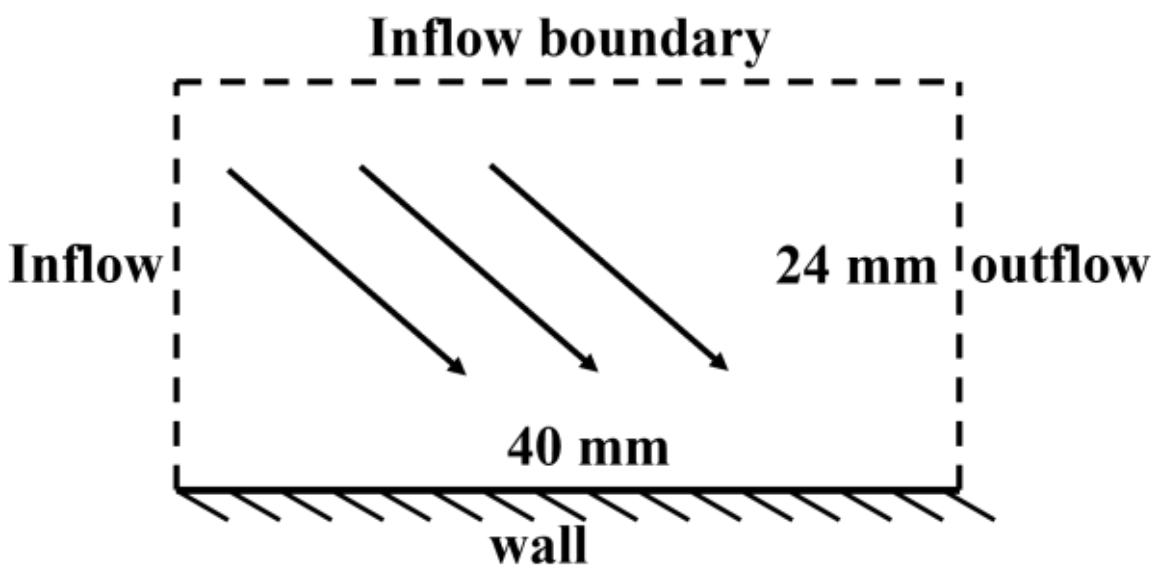

(a) Wedge-induced initiation

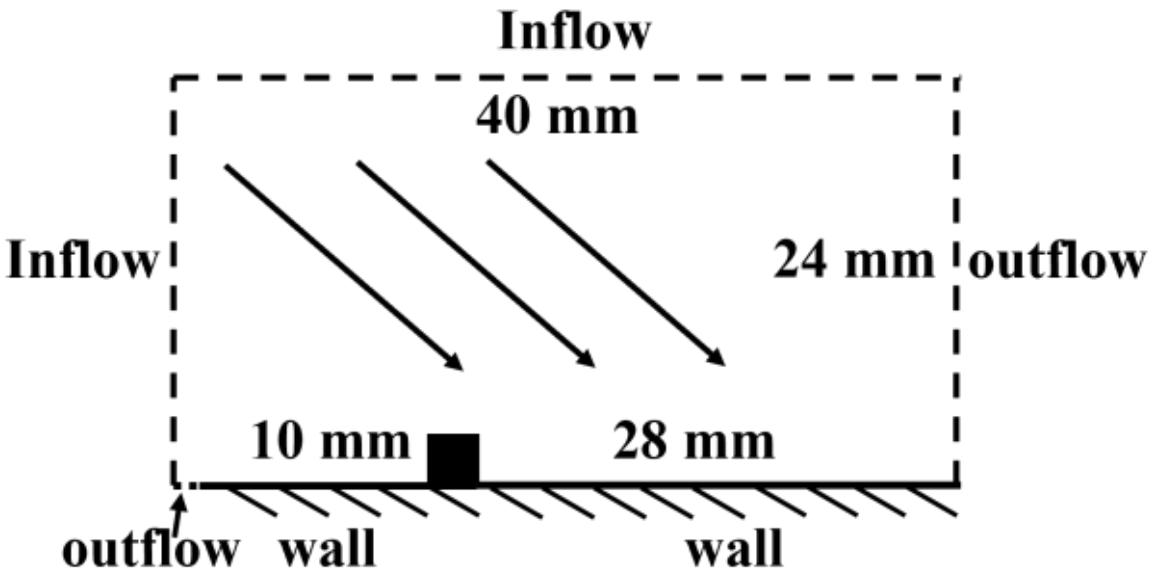

(b) Bump-forced initiation

Fig. 2. Computational domains

### 3. Results and Discussion

Numerical simulations were conducted for stoichiometric gaseous kerosene-air oblique detonation waves at a flight Mach number of 10. Table 1 summarizes the cases, and Table 2 lists the combustor-inlet conditions obtained from Ref. [23]. Because the primary objective of this study is to investigate the effect of flight altitude on oblique detonation initiation and propagation, only the combustor-inlet static pressure is varied among the cases; all other flow parameters are held constant.

**Table 1. Numerical cases**

| Cases | Parameters | Detonation status |
|---|---|---|
| 1 | 15km, 550 kPa, wedge | Yes |
| 2 | 20km, 250 kPa, wedge | Critical |
| 3 | 20km, 250 kPa, wedge + bump | Yes+ coupling |
| 4 | 30km, 50 kPa, wedge + bump | Yes+ decoupling |
| 5 | 30km, 50 kPa, large scale | Yes+ coupling |

**Table 2. The combustor-inlet conditions**

| Parameters | Values |
|---|---|
| Mach number Ma | 4.74 |
| Static Pressure P (kPa) | 550, 250, 50 |
| Static Temperature T (K) | 930 |
| Velocity u (m/s) | 2700 |

### 3.1 Cases 1 and 2

Cases 1 and 2 examine the wedge-induced initiation. Case 1 corresponds to a flight altitude of 15 km, with a combustor-inlet static pressure of 550 kPa, a static temperature of 930 K, and a mixture velocity of 2700 m/s. Case 2 corresponds to a flight altitude of 20 km and a combustor-inlet static pressure of 250 kPa.

Figure 3 shows pressure contours for Cases 1 and 2. As shown in Fig. 3(a), the oblique detonation of Case 1 is successfully initiated and propagates stably. The flow field comprises three regions: the oblique shock wave (OS) region, the initiation region, and the CJ detonation region. The oblique shock region and the initiation regions are 8 mm and 10 mm long, respectively; and they are therefore comparable in length. Transverse waves (TWs) are present on the CJ detonation front clearly. It can be found that the initiation region has a long and complex flow structure; therefore, the wedge must be sufficiently long for detonation initiation to be completed.

As shown in Fig. 3(b), increasing the flight altitude by 5 km in Case 2 halves the combustor-inlet static pressure and doubles the ignition delay time. According to the ρL binary scaling law [28-31], the lengths of the oblique shock region and initiation region also double, reaching 16 mm and 20 mm, respectively. Under the flight conditions of Case 2, the 40 mm wedge is at the critical length. Although the oblique detonation is initiated, the wedge is not long enough for the detonation wave to develop into a stable CJ detonation. The results of Cases 1 and 2 demonstrate that the oblique detonation flow field follows the ρL binary scaling law at different flight altitudes.

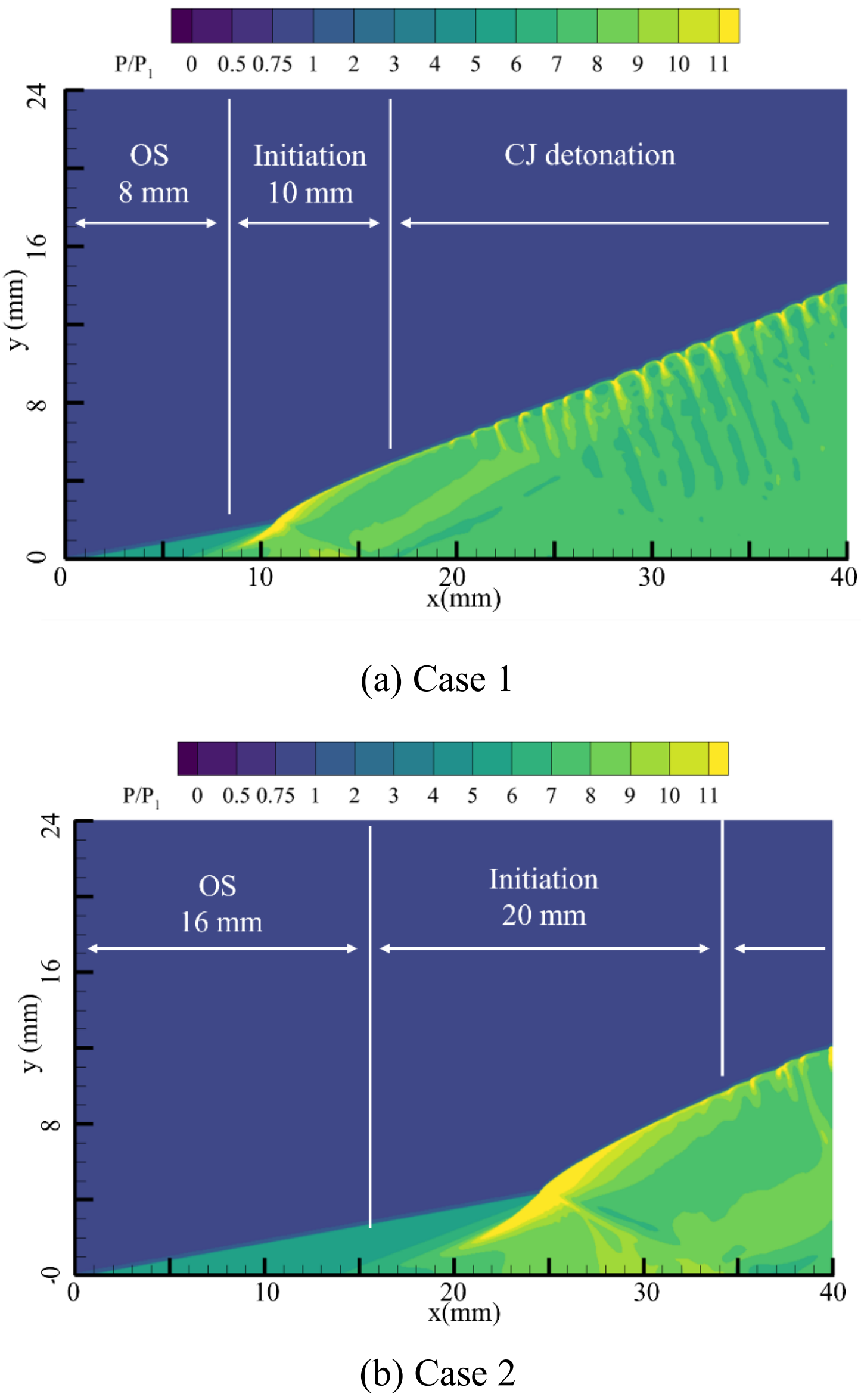

(a) Case 1

(b) Case 2

Fig. 3. Pressure contours for Cases 1 and 2

Figure 4 shows the convective flux contours (Γ) and heat release contours (Δq, black lines) over one time step for Cases 1 and 2. The convective flux analysis method is described in detail in Refs. [24-27] and is not repeated here. For Case 1, compared to transvers waves, the convective flux intensity of the oblique shock wave is very weak in the oblique shock wave region, where no chemical reaction occurs. The initiation region contains a complex convective flux structure. Once a stable CJ detonation develops, regularly spaced transverse waves propagate downstream along the oblique detonation front. Each transverse wave consists of a shock wave (blue) and a rarefaction wave (red), and its convective flux intensity is very high. Chemical heat release is tightly coupled with transverse waves, which sustain stable propagation of oblique detonation [26,27]. In Case 2, the longer ignition delay time means that, for the same wedge length, only the oblique shock wave and initiation regions are present. No distinct transverse waves appear, indicating that the oblique detonation flow field is not fully developed.

The results of Cases 1 and 2 show that the wedge-initiated oblique detonation flow field comprises an oblique shock wave region, an initiation region, and a CJ detonation region. The critical condition for obliqued detonation initiation is that the wedge length exceed the combined lengths of the oblique shock and initiation regions. Therefore, wedge-induced initiation method requires highly precise matching of the inlet-exit conditions, wedge angle and length, equivalence ratio, and other parameters. Accurate prediction and real-time control are difficult for wedge-induced initiation oblique detonation engines during maneuvering flight.

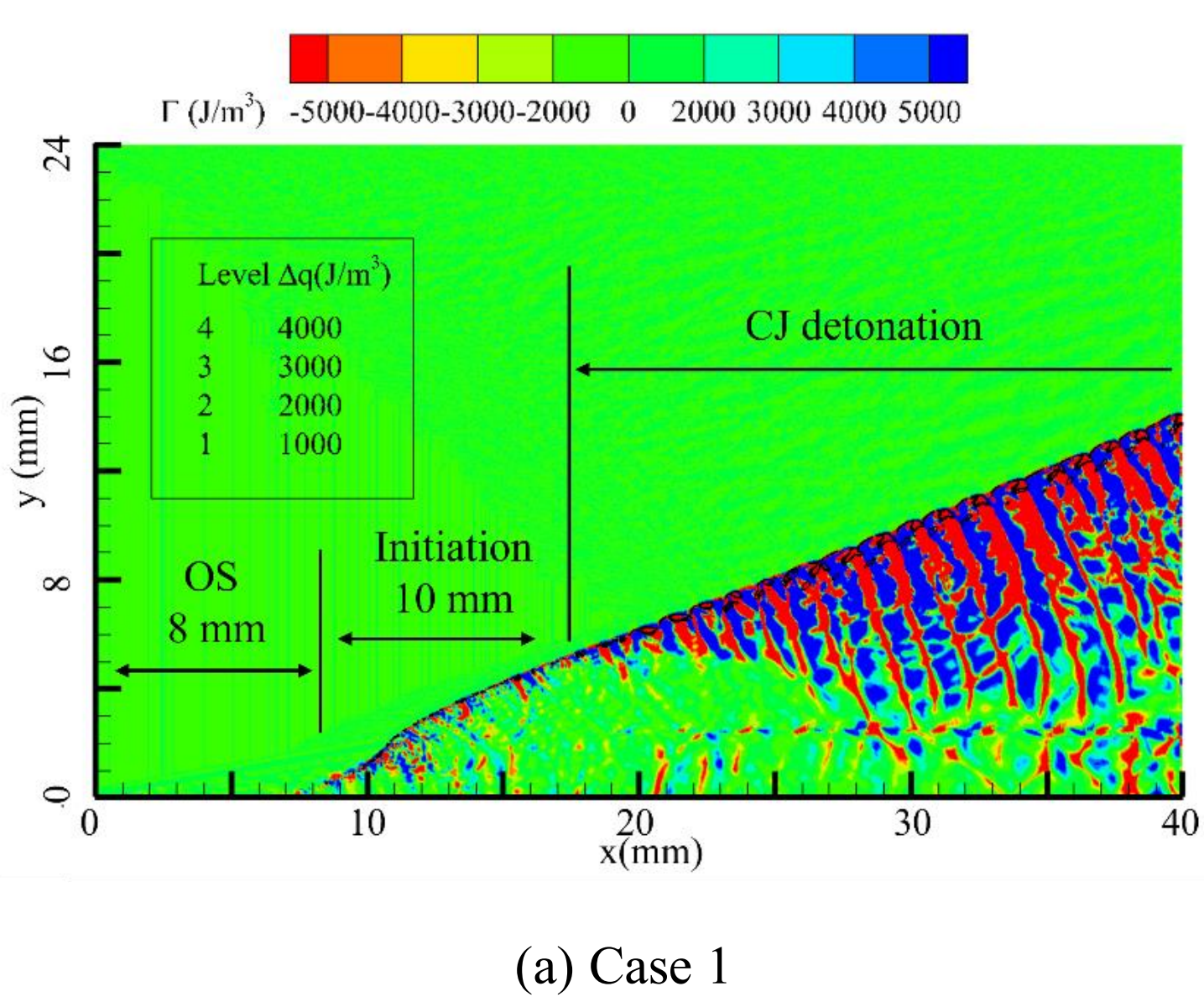

(a) Case 1

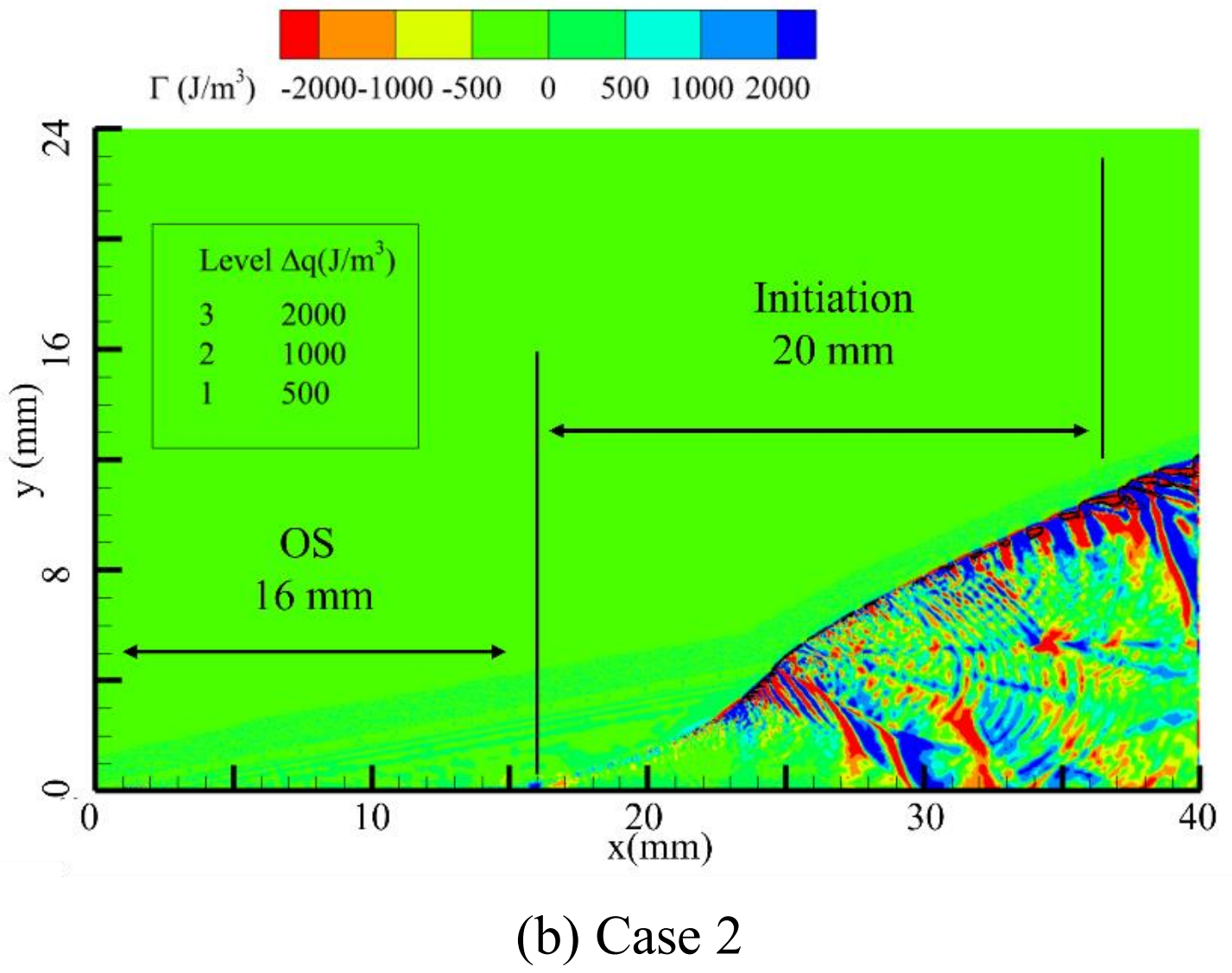

(b) Case 2

Fig. 4. Convective flux and heat release for Cases 1 and 2

### 3.2 Cases 3 and 4

Cases 3 and 4 examine the bump-forced initiation at flight altitudes of 20 and 30 km, respectively, with all other parameters held constant. Figure 5 shows pressure contours and β contours (white lines) for these two cases, and Fig. 6 compares their pressure contours. It can be seen that after forced initiation by the bump, a stable oblique detonation wave forms, and its initiation location is independent of flight altitudes because the bump-forced initiation exploits the high total temperature and total pressure at the stagnation point to ignite the mixture.

Different from the wedge-induced initiation, the bump generates an expansion wave that interacts with the oblique detonation wave downstream. As shown in Fig. 6, the upstream region is unaffected by the expansion wave, and these two cases therefore have identical upstream flow structures. Downstream, the expansion wave slows the chemical reaction rate and increases the ignition delay time. At the flight altitude of 20 km, as shown in Fig. 5(a), the expansion wave does not immediately decouple the oblique detonation. Transverse waves form and couple the chemical heat release to the oblique shock, allowing the oblique detonation to continue propagating stably [26,27]. However, at higher altitude of 30 km, as shown in Fig. 5(b), the lower pressure and longer ignition delay time prevent transverse waves from forming under the influence of expansion waves, and the oblique detonation gradually decouples downstream.

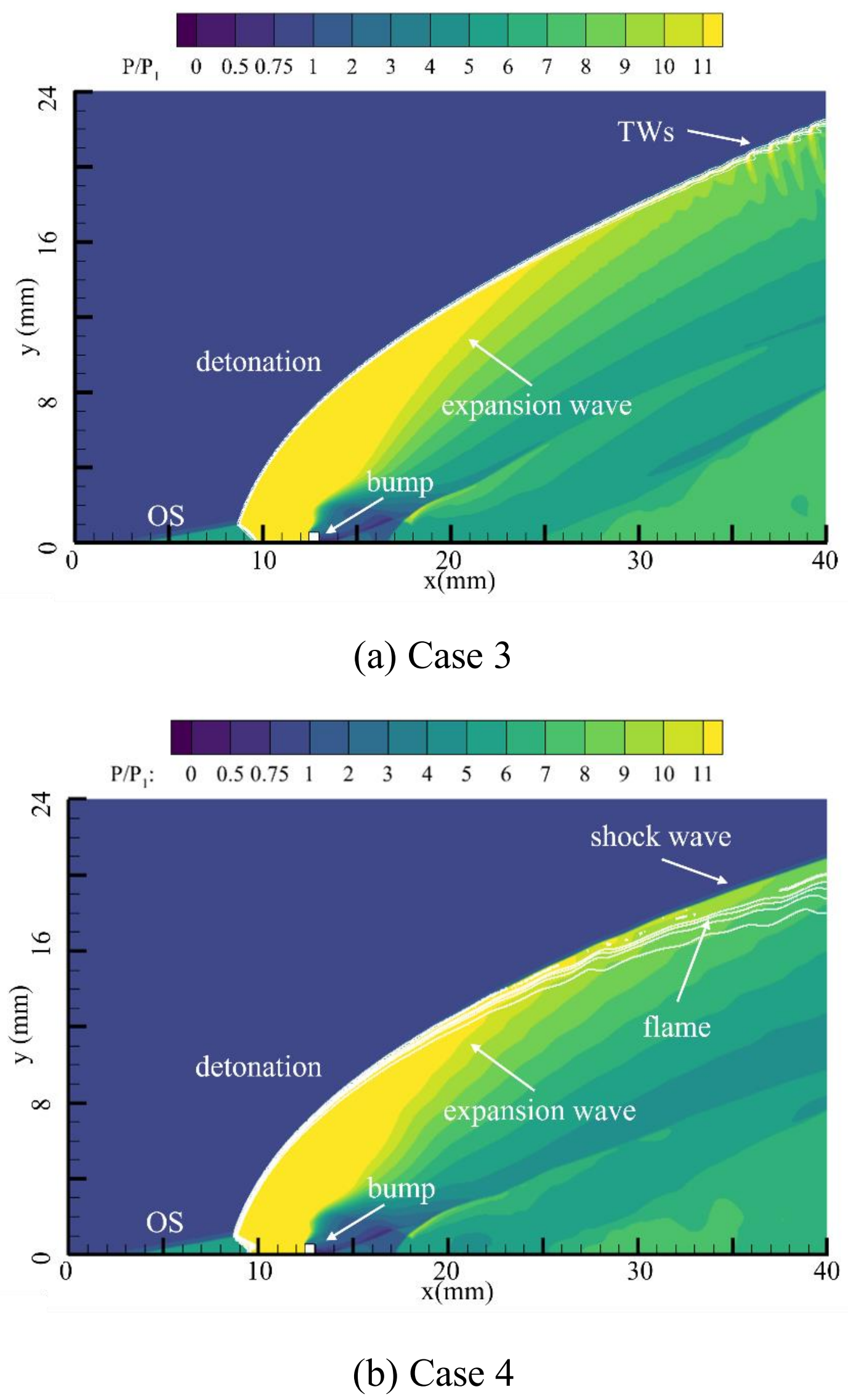

(a) Case 3

(b) Case 4

Fig. 5. Pressure and β contours for Cases 3 and 4

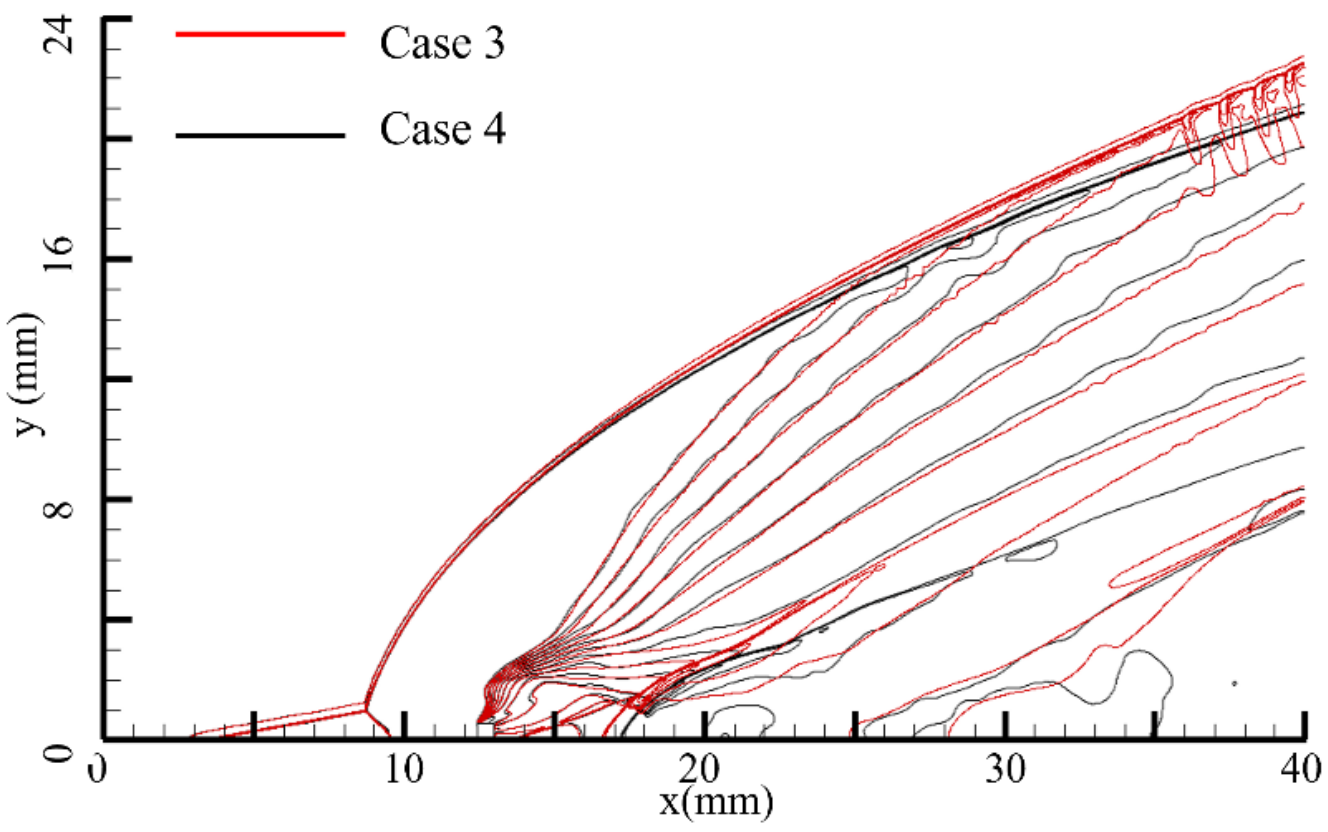

Fig. 6. Comparison of pressure contours for Cases 3 and 4

Figure 7 shows convective flux and heat release contours for these two cases. Figure 8 shows contours of the pressure gain over one time step (Δp, white lines), whose definition is provided in Refs. [24-27]. The pressure gain over one time step results from the combined effects of convection and chemical heat release and depends on whether the heat release is coupled with shock wave or rarefaction wave. The convective flux is positive for shock wave and negative for rarefaction wave. If the algebraic sum of the convective flux and chemical heat release is positive, the chemical heat release produces a pressure gain; if the sum is negative, no pressure gain is produced.

As shown in Fig. 7, transverse waves are present clearly on the oblique detonation front at the flight altitude of 20 km, tightly coupling chemical heat release with the shock waves. Their appearance indicates that the oblique detonation has become unstable but has not yet decoupled. Transverse waves are a propagation stage preceding oblique detonation decoupling. At 30 km, no transverse wave appears, and the flame gradually decouples from the oblique shock. Figure 8 shows that, at 20 km, coupling between chemical heat release and the transverse waves produces a pressure gain (the white lines and black lines overlap), thereby sustaining stable propagation. At 30 km, the absence of transverse waves causes the chemical heat release to couple with rarefaction wave (black lines and red contours overlap). Consequently, the heat release produces no pressure gain (no white lines), and the oblique detonation wave gradually decouples downstream.

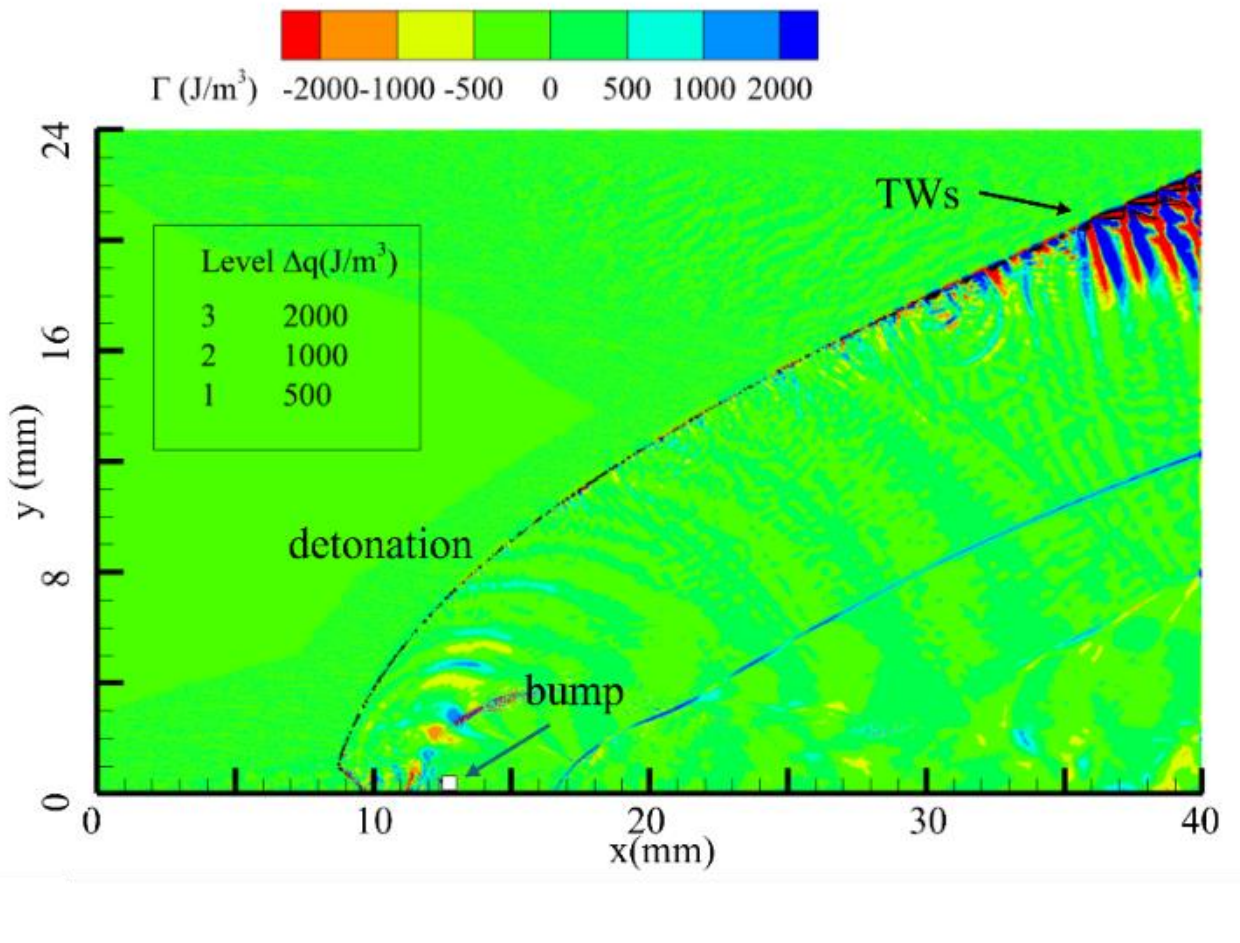

(a) Case 3

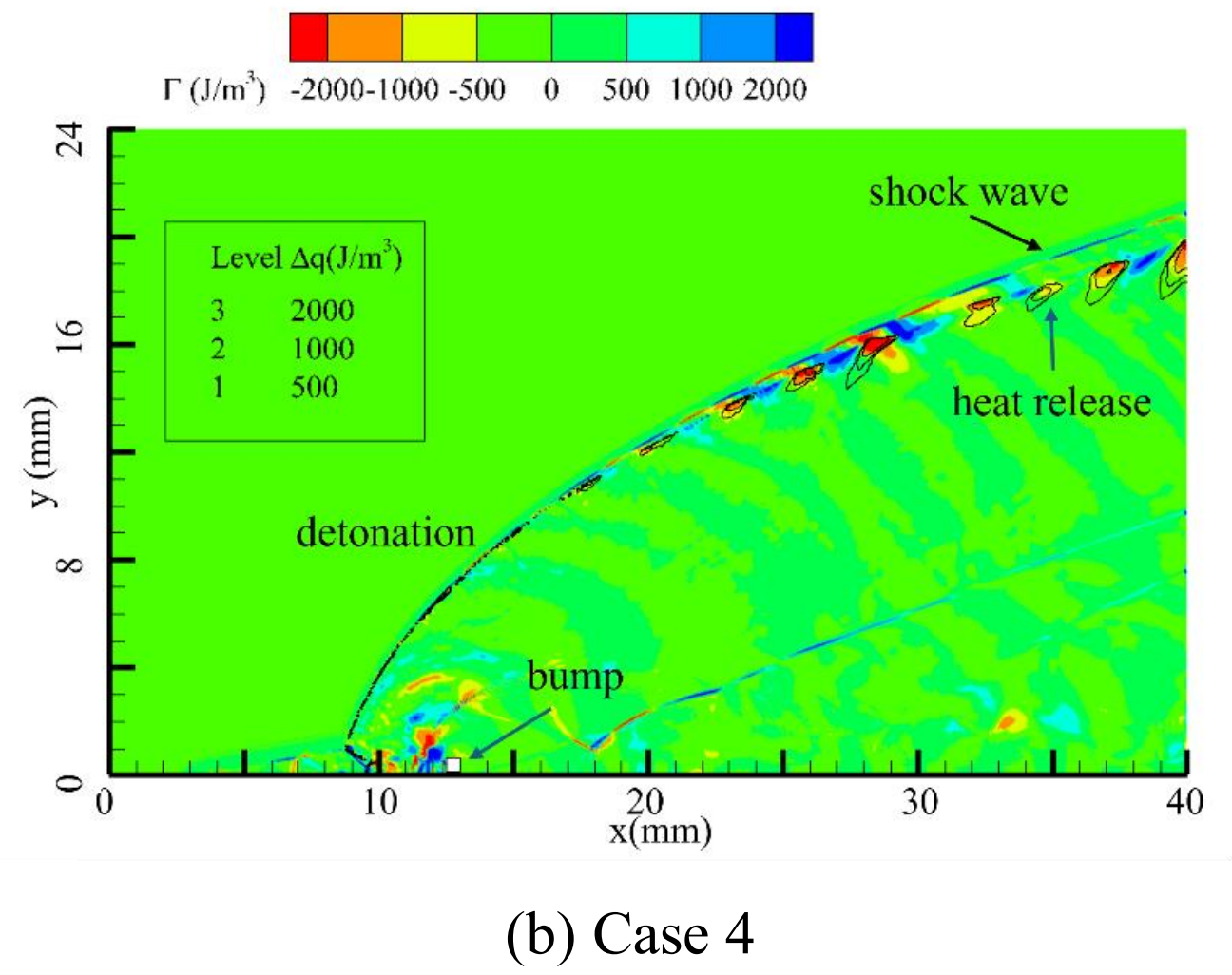

(b) Case 4

Fig. 7. Convective flux and heat release contours for Cases 3 and 4

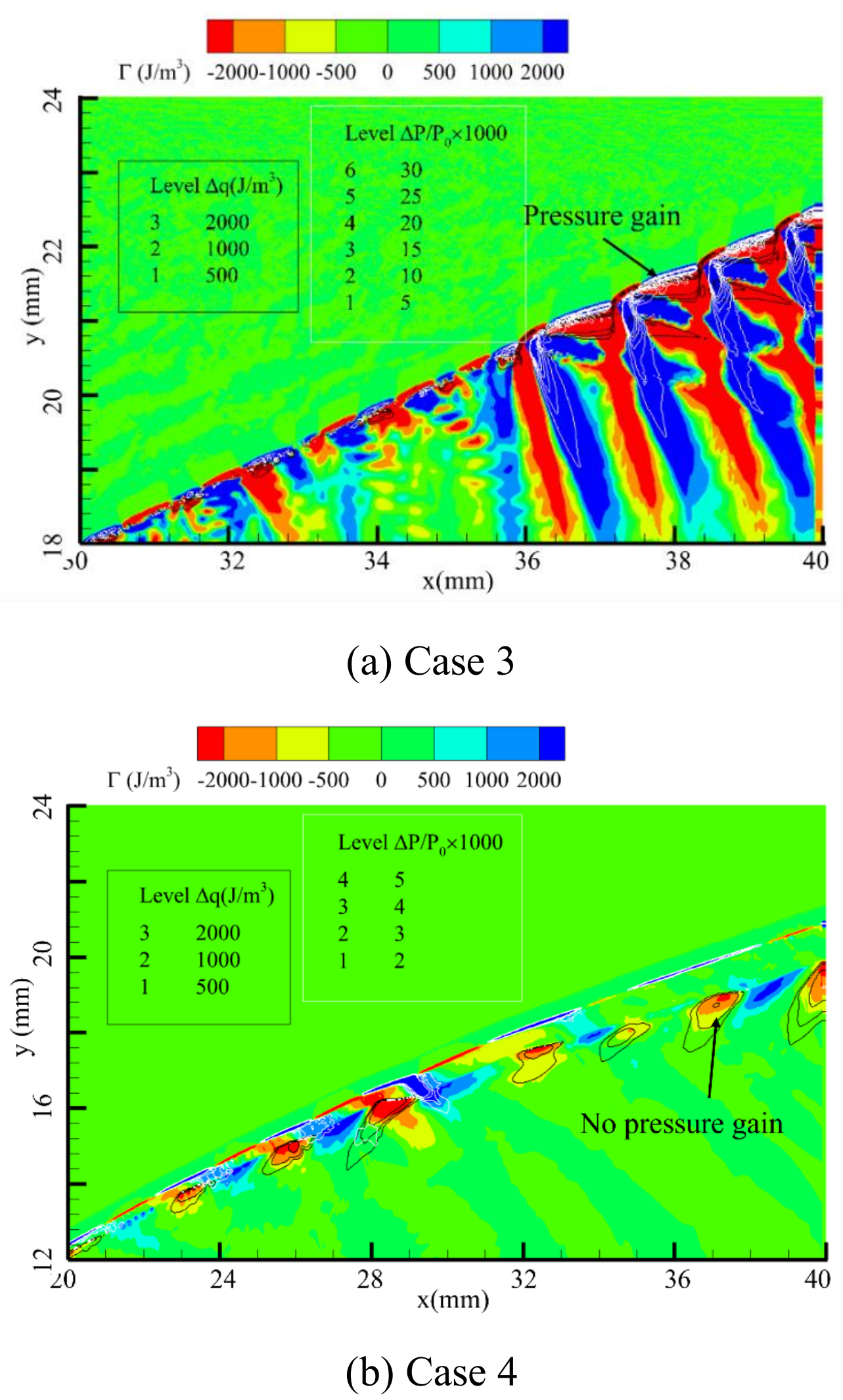

(a) Case 3

(b) Case 4

Fig. 8. Enlarged local contours of pressure gain, convective flux, and heat release for Cases 3 and 4

The results of Cases 3 and 4 show that bump-forced initiation exploits the high total temperature and total pressure at the stagnation point of bump to ignite the mixture and does not require any precise matching of the inlet-exit conditions, wedge angle and length, equivalence ratio, or other parameters. At different flight altitudes, the upstream flow fields of oblique detonation are identical; gradual decoupling occurs only downstream under the influence of the expansion wave. Thus, the oblique detonation waves established by bump-forced initiation is predictable before a flight test and controllable during maneuver flight.

### 3.3 Case 5

Case 5 is used to validate the ρL binary scaling law. Its combustor-inlet conditions are identical to Case 4, but its computational domain is five times larger. According to the ρL binary scaling law, the flow field of Case 5 should be identical to that of Case 3. To reduce the computational cost while keeping the total number of grid cells unchanged, the grid spacing is increased from 20 μm to 100 μm. Figure 9 shows the pressure and convective flux contours for this case. Transverse waves appear evidently on the oblique detonation front, and the oblique detonation wave does not decouple downstream. Figure 10 compares the pressure contours of Cases 3 and 5. Their flow fields are identical, confirming that the oblique detonation wave follows the ρL binary scaling law.

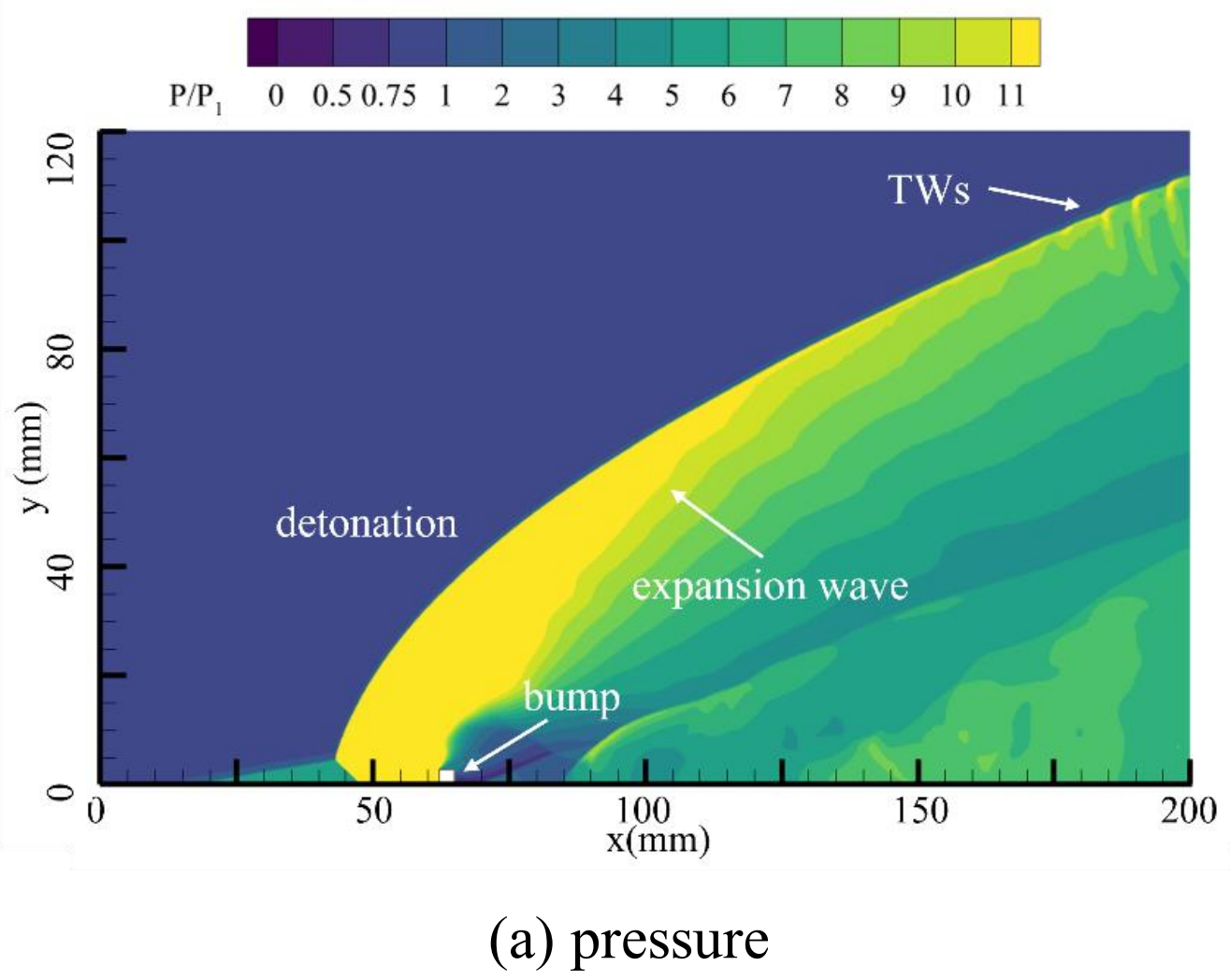

(a) pressure

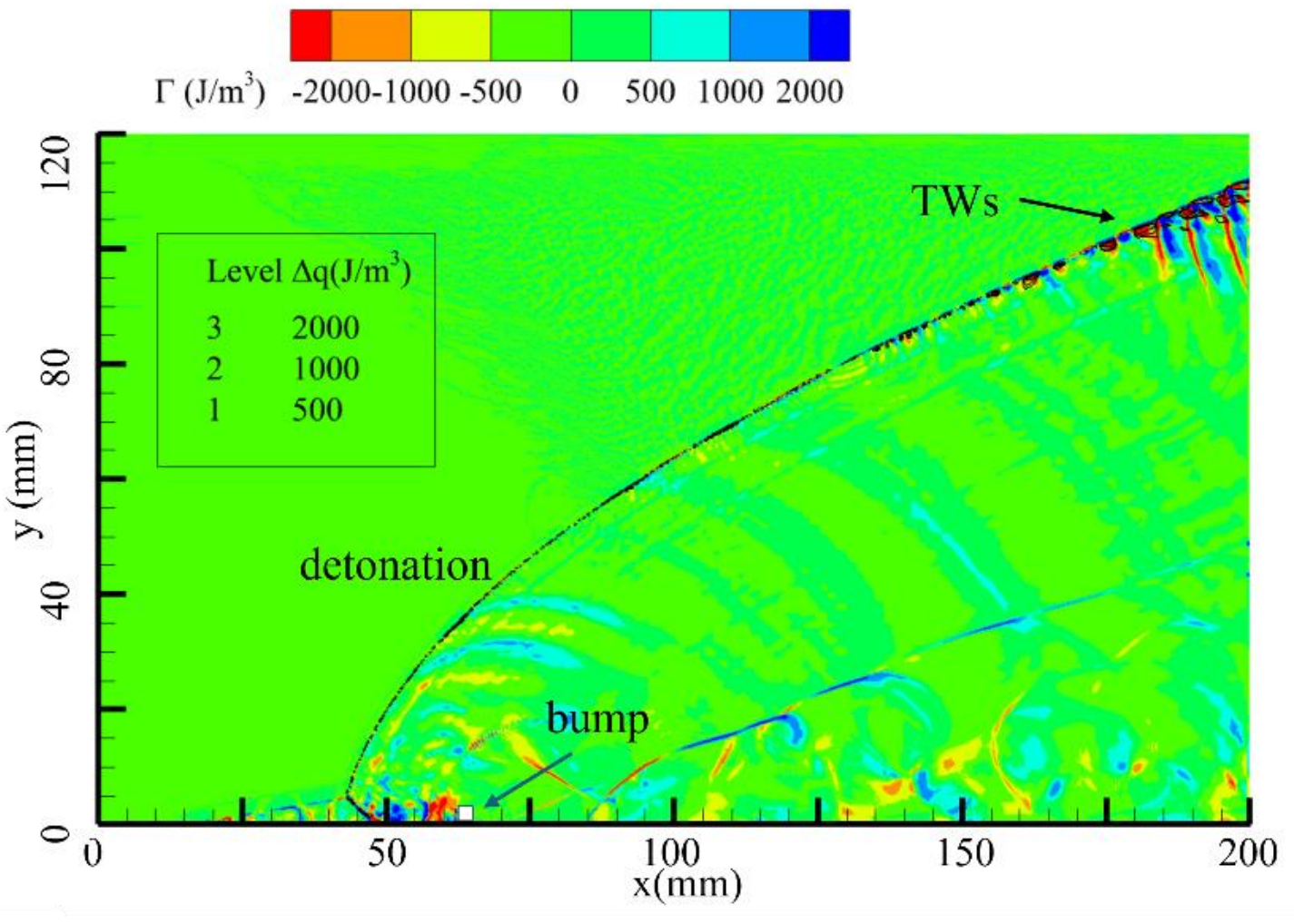

(b) convective flux and heat release

Fig. 9. Pressure and convective flux contours for Case 5

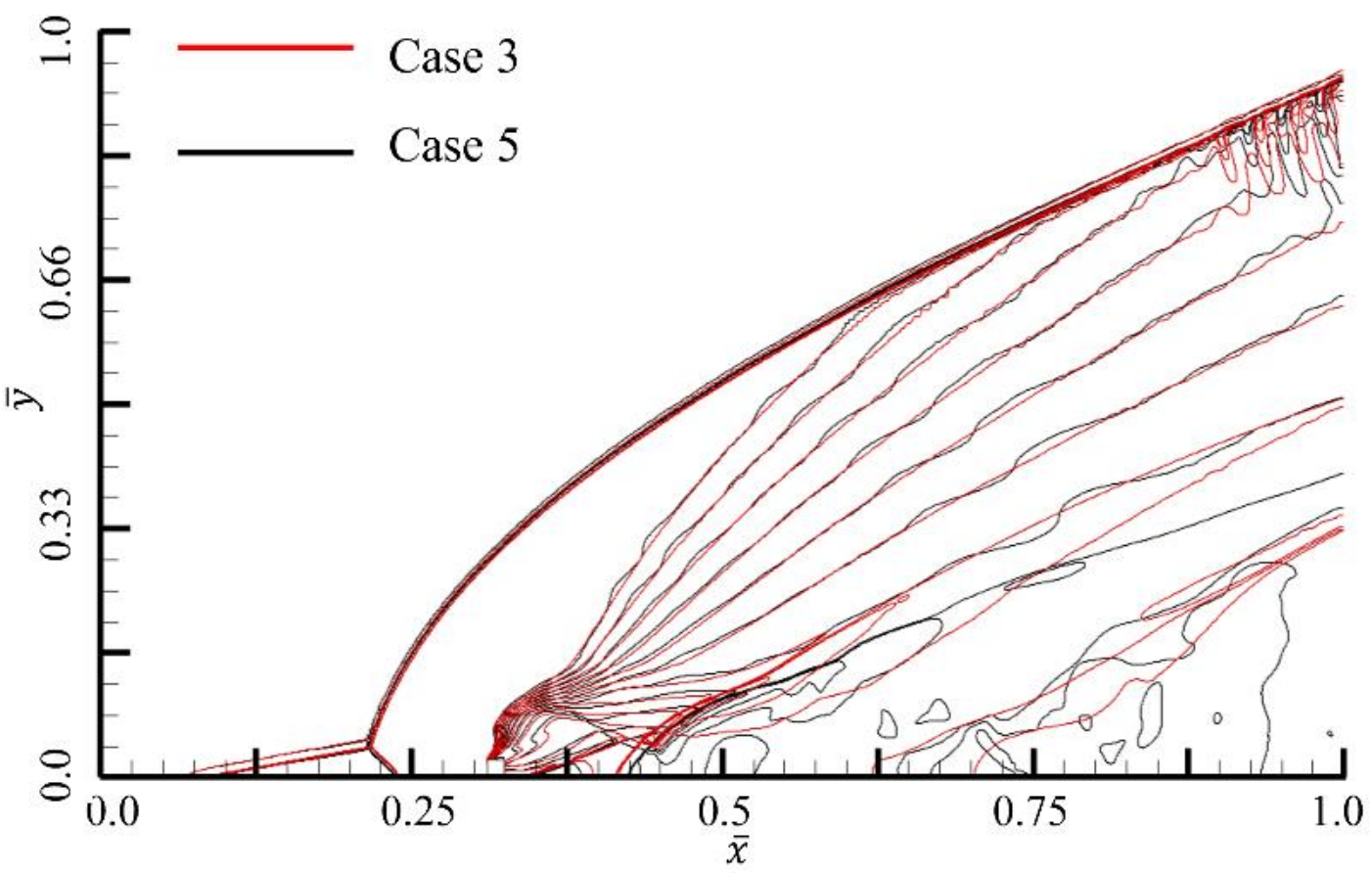

Fig. 10. Comparison of pressure contours for Cases 3 and 5

## 4. Validation

The numerical method and chemical reaction model were validated against shock-tunnel free-jet experimental results for a kerosene-air oblique detonation [10,11]. The computational domain is identical to that shown in Fig. 2(b). The experiment simulated a flight Mach number of 9 at an altitude of 45 km and employed bump-forced initiation, a wedge angle of 30°, and a mixture equivalence ratio of 1.0. Table 3 lists the combustor-inlet conditions after inlet compression. In the experimental model, the bump is 50 mm downstream of the wedge tip. In Case A, the bump is 10 mm downstream of the wedge tip, and the computational domain is a 1:5 scaled version of the experimental model. The experimental static pressure is 5 kPa. According to the ρL binary scaling law, the static pressure in Case A is therefore set to 25 kPa, while all the other parameters remain unchanged.

**Table 3. Combustor-inlet conditions for bump-forced kerosene-air oblique detonation experiments**

| Parameters | Experiments | Case A | Case B |
|---|---|---|---|
| Wedge angle, θ (°) | 30 | 30 | 30 |
| Grid spacing (μm) | | 20 | 100 |
| Bump size (mm) | 5 | 0.6 | 3 |
| Length from wedge tip to bump (mm) | 50 | 10 | 50 |
| Mach number, Ma | 4.5 | 4.5 | 4.5 |
| Static pressure, P (kPa) | 5 | 25 | 5 |
| Static temperature, T (K) | 800 | 800 | 800 |
| Velocity, u (m/s) | 2400 | 2400 | 2400 |

Figure 11 compares the experimental schlieren image with the Case A results. Figure 11(a) shows pressure and β contours (black lines). Following forced initiation by the bump, the overdriven oblique detonation wave moves upstream along the wedge surface to the wedge tip and remains stable. The expansion wave generated by the bump interacts with the oblique detonation downstream. The oblique detonation remains coupled in the upstream region, which is unaffected by the expansion wave, and gradually decouples downstream under its influence. The experimental schlieren image in Fig. 11(b) clearly shows that the oblique detonation wave is stabilized at the wedge tip and decouples gradually downstream. Figures 11(c) and 11(d) compare the schlieren image with the pressure and β contours, respectively. The numerical results agree very closely with the experimental results, validating the numerical method and chemical reaction model.

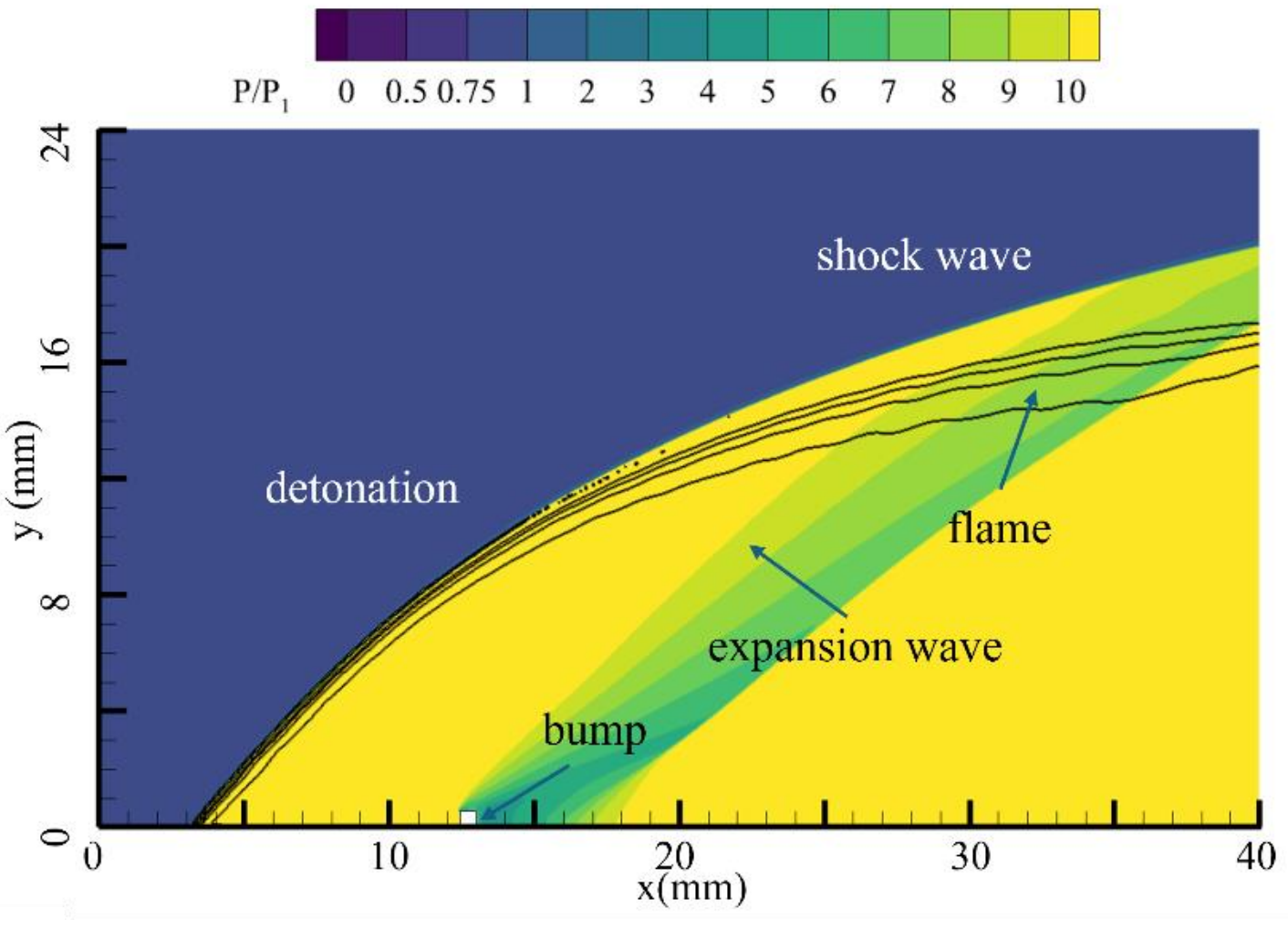

(a) Pressure and β contours

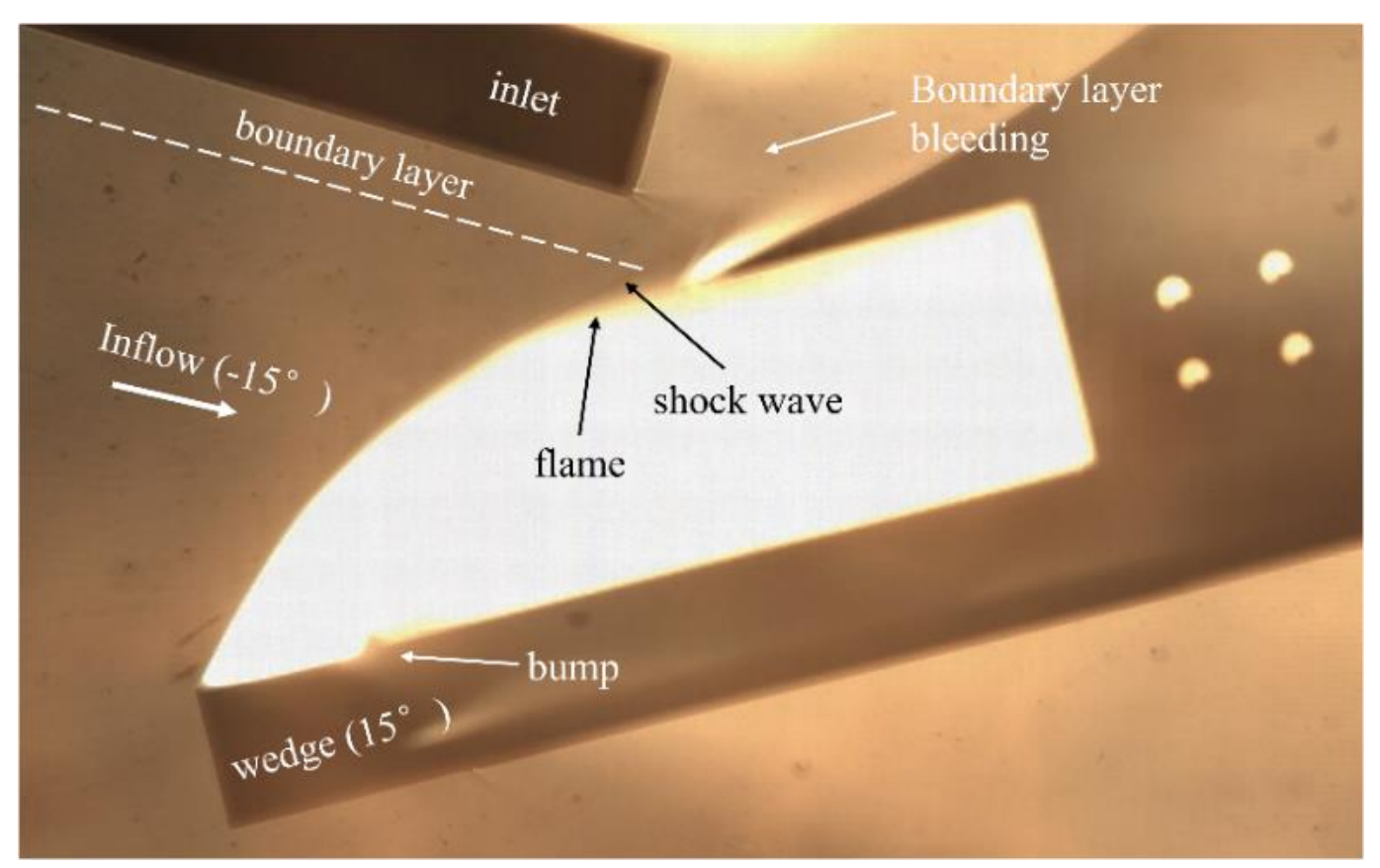

(b) Experimental schlieren image

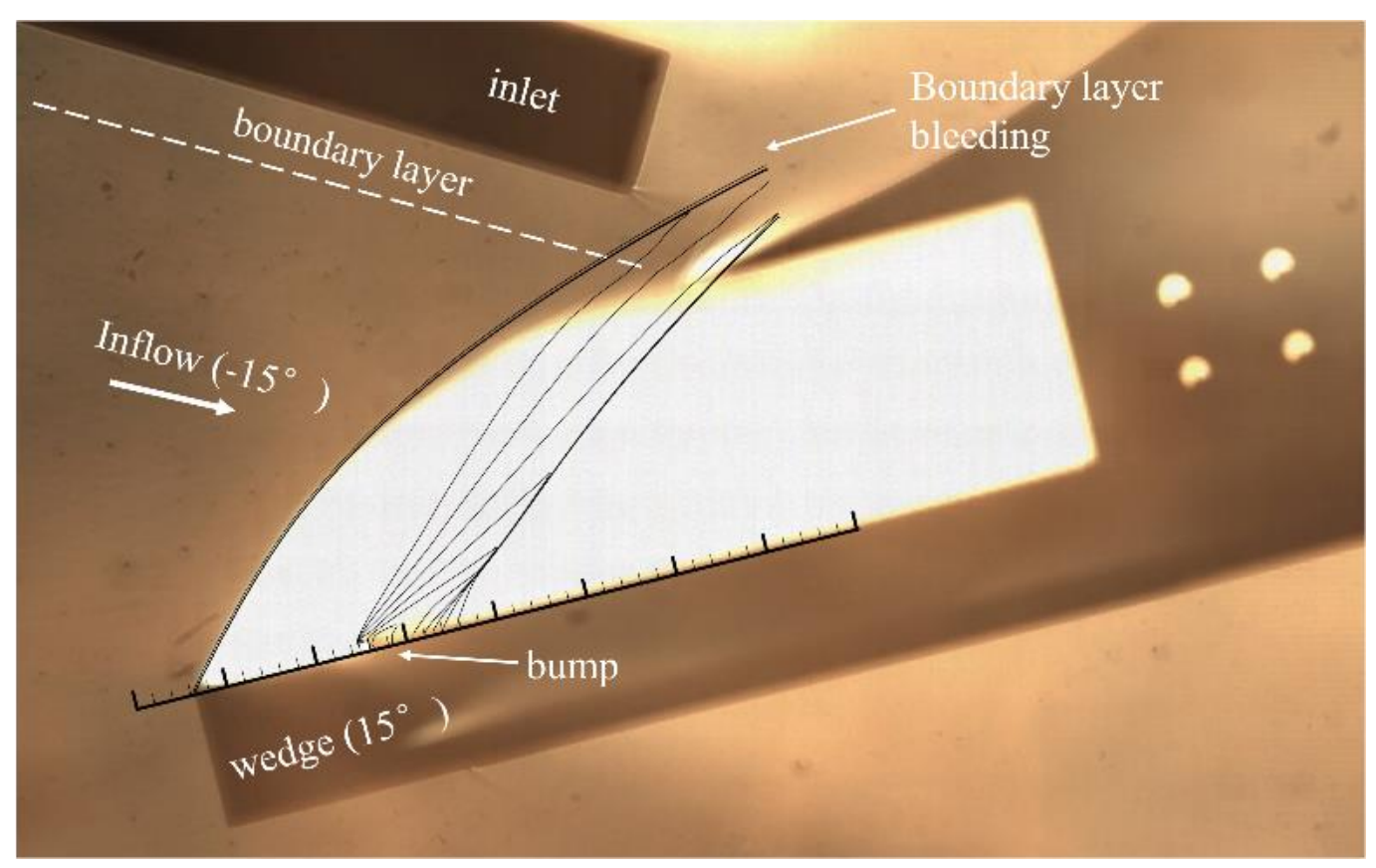

(c) Schlieren image and pressure contours

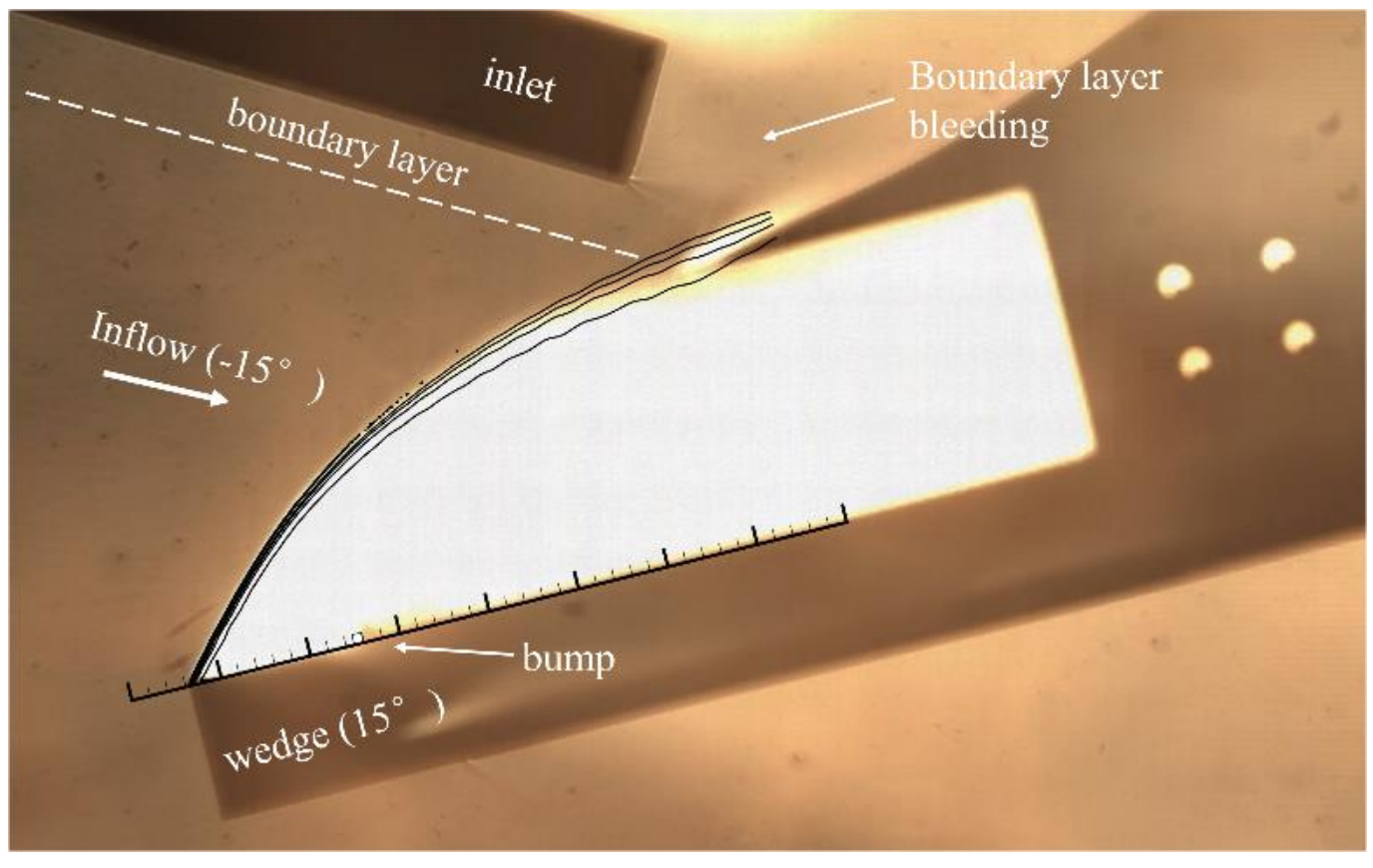

(d) Schlieren image and β contours

Fig. 11. Comparison of the experimental schlieren image and Case A

In Case B, the grid spacing is increased from 20 μm to 100 μm, while all other parameters remain unchanged. The computational domain has a 1:1 scale relative to the experimental model, the bump is located 50 mm downstream of the wedge tip, and the static pressure is set to be 5 kPa, being the same as experiments. Figure 12 compares the pressure and β contours of Cases A and B. The pressure contours of these two cases agree closely, further demonstrating that the flow field of oblique detonation wave follows the ρL binary scaling law.

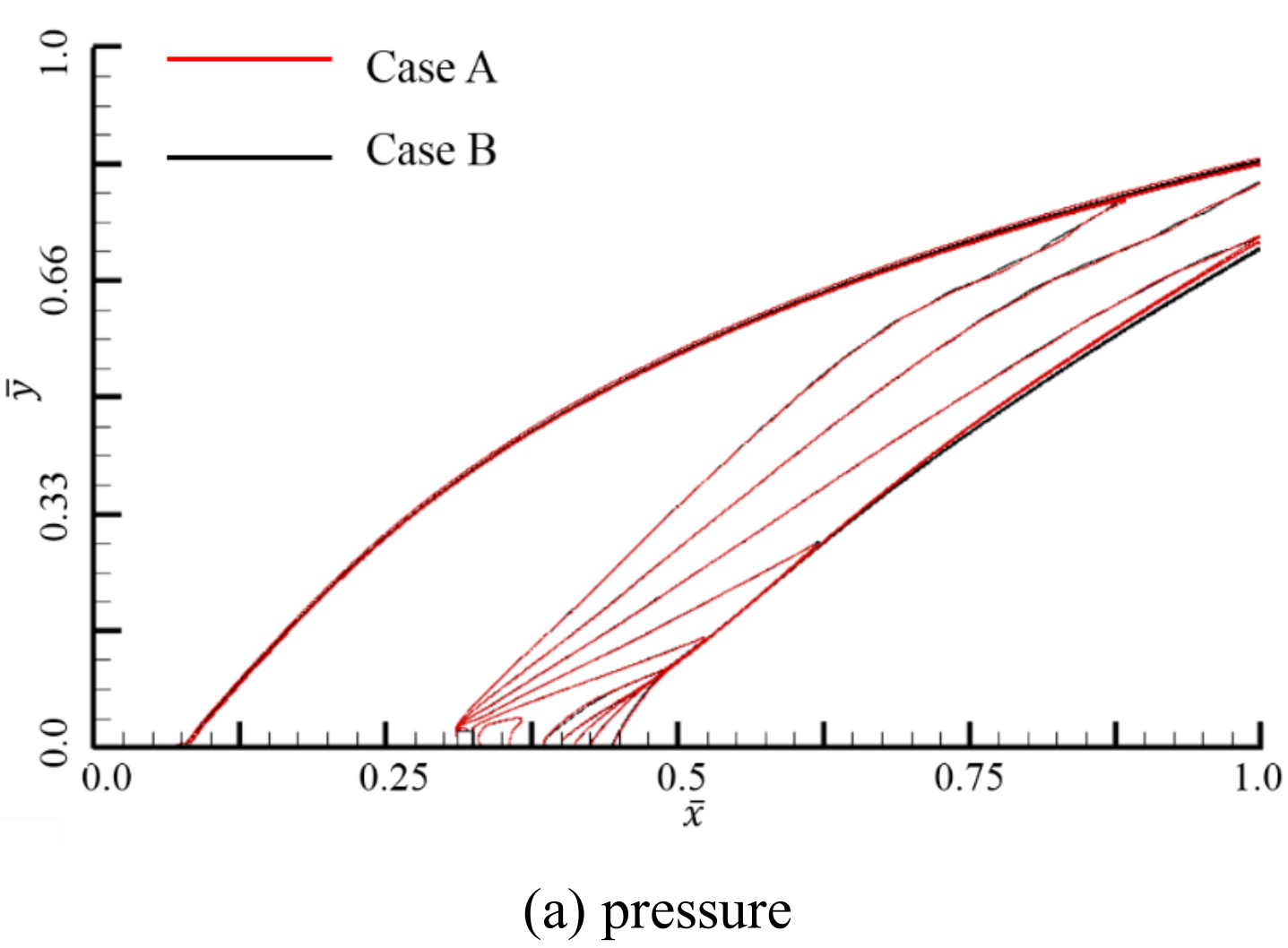

(a) pressure

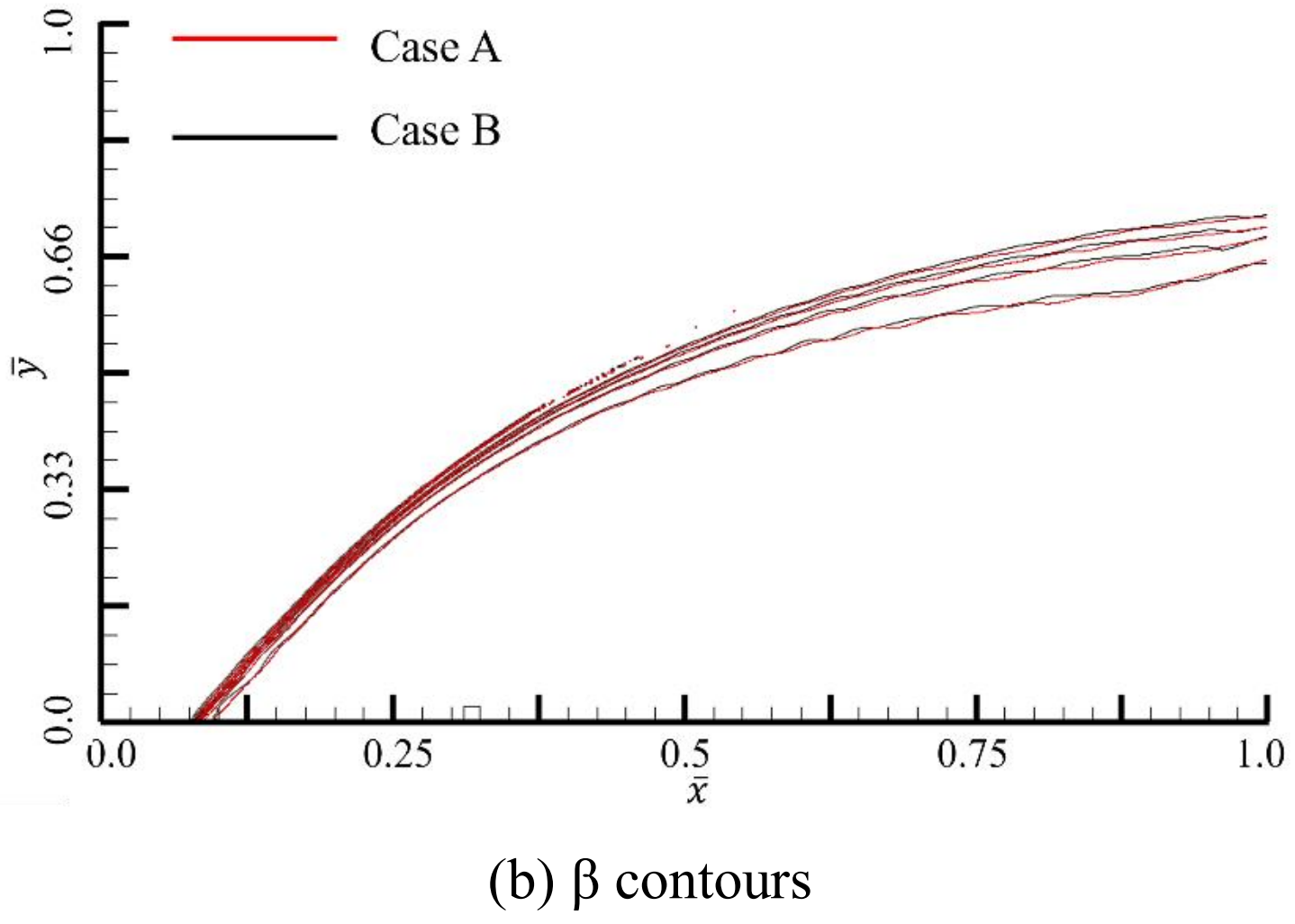

(b) β contours

Fig. 12. Comparison of pressure and β contours for Cases A and B

Additionally, Li et al. also conducted two-dimensional numerical simulations of kerosene-fueled oblique detonation waves using a modified solver based on OpenFOAM [32]. A two-step mechanism of kerosene has been employed [33]. The flow conditions are the same as Case 1. In their study, three different mesh sizes of 30 μm, 40 μm, and 50 μm were analyzed and the size of 40 μm was selected for numerical simulations. The pressure contours of Case 1 are compared with theirs in Fig.13. It can be seen that transverse waves appear clearly on the detonation front, the oblique denotation wave angle is 42°, and there is a good agreement between them.

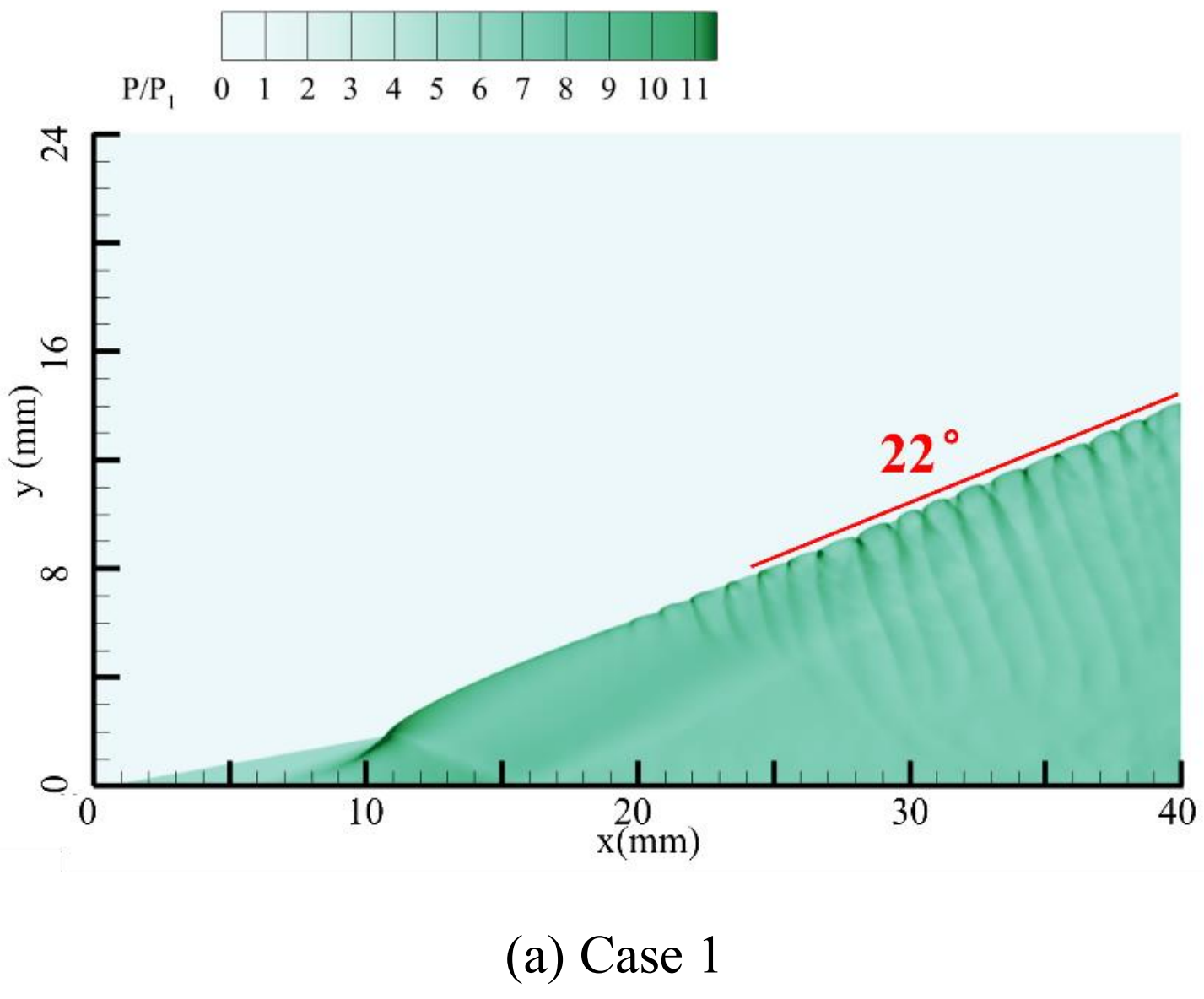

(a) Case 1

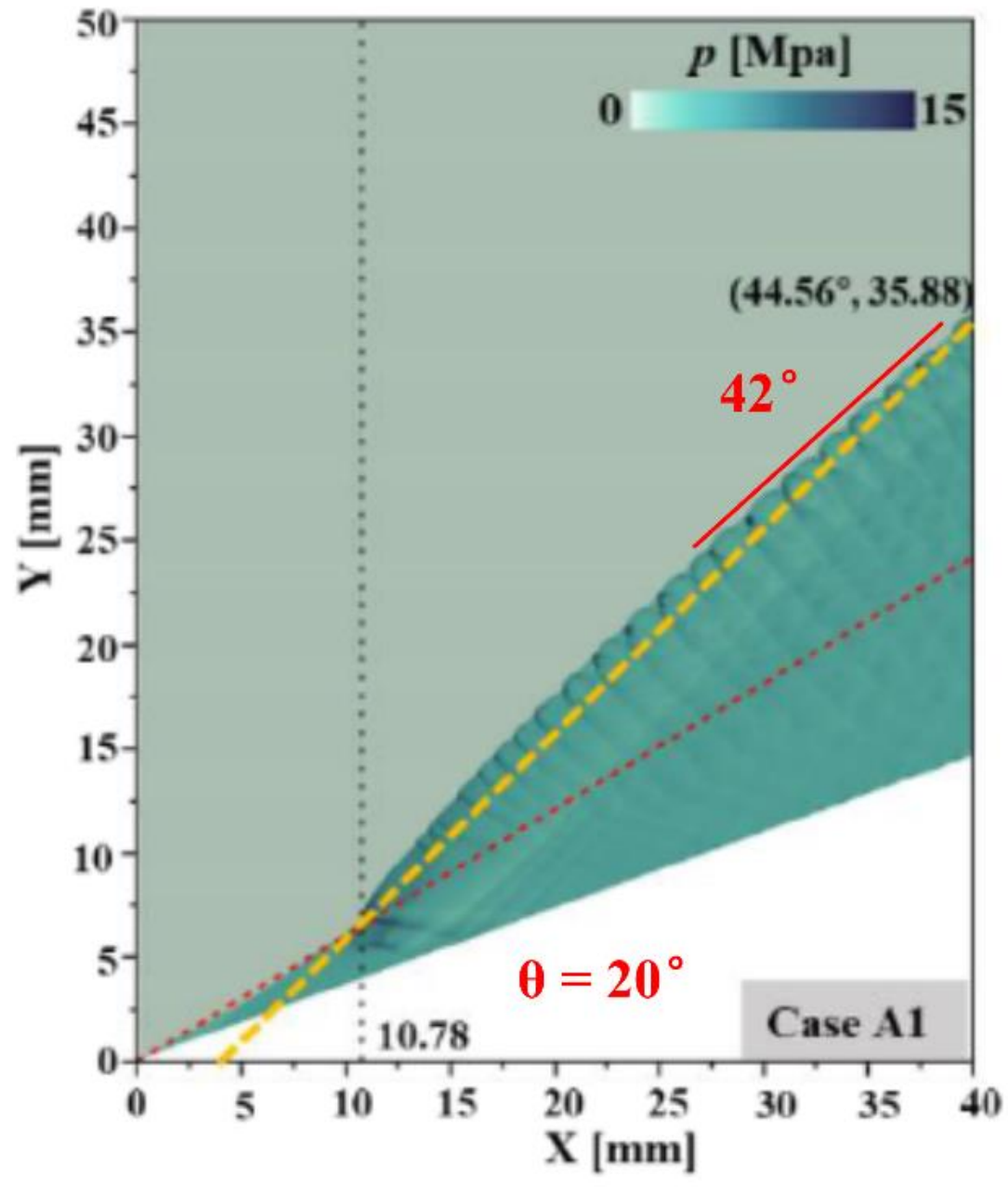

(b) the Reference [32]

Fig. 13. Comparison of oblique detonation angle with Refence [32]

## 5. Conclusions

Numerical simulations of Mach 10 kerosene-fueled oblique detonation waves at different flight altitudes were conducted using the two-dimensional conservative Euler equations and a second-order two-step global chemical reaction model. The objective was to compare wedge-induced initiation with bump-forced initiation. The main conclusions are as follows:

(1) At a fixed flight Mach number and configuration, neglecting the temperature variation with altitudes, the oblique detonation flow field at different flight altitudes follows the ρL binary scaling law.

(2) For wedge-induced initiation, the oblique detonation flow field comprises an oblique shock wave region, an initiation region, and a CJ detonation region. The critical wedge length must exceed the combined lengths of the oblique shock wave and initiation regions. Wedge-induced initiation therefore requires highly precise matching of the inlet-exit conditions, wedge angle and length, equivalence ratio, and other parameters.

(3) Bump-forced initiation exploits the high total temperature and total pressure at the stagnation point; therefore, the initiation location is fixed. It does not require precise matching of the inlet-exit conditions, wedge angle and length, equivalence ratio, or other parameters. Consequently, bump-forced initiation is highly reliable and produces a stable oblique detonation during flight.

(4) The bump generates an expansion wave that interacts with the oblique detonation downstream and slows the chemical reaction rate. At lower flight altitude, transverse waves appear and the oblique detonation keeps stable propagation in the expansion wave influence region. As flight altitude increases, the oblique detonation wave gradually decouples downstream.